\documentclass{osa-article}

\journal{oe}

\usepackage{graphicx}
\usepackage{dcolumn}
\usepackage{bm}
\usepackage{hyperref}
\usepackage{algorithm}
\usepackage{algpseudocode}
\usepackage{siunitx}
\usepackage{listings}
\usepackage{color,soul}
\usepackage{subcaption}

\lstdefinelanguage{ini} {
    basicstyle=\ttfamily\scriptsize,
    columns=fullflexible,
    morecomment=[s][\color{red}\bfseries]{[}{]},
    morecomment=[l]{\#},
    commentstyle=\color{blue}\ttfamily,
    morekeywords={},
    otherkeywords={=,:::},
    keywordstyle={\color{blue}\bfseries}
}

\begin{document}

\title{Low-signal limit of X-ray single particle diffractive imaging}

\author{Kartik Ayyer,\authormark{1,2,*} Andrew J. Morgan,\authormark{2,13} Andrew A. Aquila,\authormark{3} Hasan DeMirci,\authormark{4,5,6} Brenda G. Hogue,\authormark{7,8,9} Richard A. Kirian,\authormark{10} P. Lourdu Xavier,\authormark{1,2,3} Chun Hong Yoon,\authormark{3} Henry N. Chapman,\authormark{2,11,12} and Anton Barty\authormark{2}}

\address{\authormark{1}Max Planck Institute for the Structure and Dynamics of Matter, Luruper Chaussee 149, 22761, Hamburg, Germany\\
\authormark{2}Center for Free-Electron Laser Science, Deutsches Elektronen Synchrotron DESY, Notkestra{\ss}e 85, 22607 Hamburg, Germany\\
\authormark{3}Linac Coherent Light Source, SLAC National Accelerator Laboratory, 2575 Sand Hill Road, Menlo Park, CA, 94025, USA\\
\authormark{4}Biosciences Division, SLAC National Accelerator Laboratory, 2575 Sand Hill Road, Menlo Park, CA, 94025, USA\\
\authormark{5}Stanford PULSE Institute, SLAC National Accelerator Laboratory, 2575 Sand Hill Road, Menlo Park, CA, 94025, USA\\
\authormark{6}Department of Molecular Biology and Genetics, Koc University, Rumelifeneri yolu, Sariyer, Istanbul, 34450 Turkey\\
\authormark{7}Biodesign Center for Immunotherapy, Vaccines, and Virotherapy, Biodesign Institute at Arizona State University, Tempe 85288, USA\\
\authormark{8}Biodesign Center for Applied Structural Discovery, Biodesign Institute at Arizona State University, Tempe 85287, USA\\
\authormark{9}Arizona State University, School of Life Sciences (SOLS), Tempe, Arizona 85287, USA\\
\authormark{10}Department of Physics, Arizona State University, Tempe, AZ 85287, USA\\
\authormark{11}Department of Physics, Universit{\"a}t Hamburg, Luruper Chaussee 149, Hamburg, Germany\\
\authormark{12}The Hamburg Center for Ultrafast Imaging, Universit{\"a}t Hamburg, Luruper Chaussee 149, Hamburg, Germany\\
\authormark{13}Currently with the ARC Centre of Excellence for Advanced Molecular Imaging, School of Physics, The University of Melbourne, Parkville, VIC 3010, Australia}

\email{\authormark{*}kartik.ayyer@mpsd.mpg.de} 



\begin{abstract}
An outstanding question in X-ray single particle imaging experiments has been the feasibility of imaging sub 10-nm-sized biomolecules under realistic experimental conditions where very few photons are expected to be measured in a single snapshot and instrument background may be significant relative to particle scattering. While analyses of simulated data have shown that the determination of an average image should be feasible using Bayesian methods such as the EMC algorithm, this has yet to be demonstrated using experimental data containing realistic non-isotropic instrument background, sample variability and other experimental factors. In this work, we show that the orientation and phase retrieval steps work at photon counts diluted to the signal levels one expects from smaller molecules or with weaker pulses, using data from experimental measurements of 60-nm PR772 viruses. Even when the signal is reduced to a fraction as little as 1/256, the virus electron density determined using \emph{ab initio} phasing is of almost the same quality as the high-signal data. However, we are still limited by the total number of patterns collected, which may soon be mitigated by the advent of high repetition-rate sources like the European XFEL and LCLS-II.
\end{abstract}

\section{\label{sec:intro}Introduction}
The potential of X-ray free electron lasers (XFELs) to image biomolecular structures at room temperature without the need for crystallisation has been one of the goals driving their development. For many years, theoretical studies backed by simulated data have suggested that near-atomic resolution of isolated non-crystalline proteins should be possible with currently available XFEL sources~\cite{Neutze:2000, Ayyer:2016}. To date, published results have focused on large or symmetric particles such as viruses in the 60-500nm size range where the higher signal levels from larger particles is ideal for methods development \cite{Loh:2010,Kassemeyer:2013,Ekeberg:2015,Aquila:2018}. Results from the single particle imaging initiative at the Linac Coherent Light Source (LCLS) \cite{Aquila:2015} have been in a similar size range~\cite{Munke:2016,Reddy:2017}. 

Imaging individual proteins has so far proven more elusive due to the lower signal-to-background from smaller sized particles and a lower than expected rate of single particle diffraction pattern acquisition~\cite{Aquila:2018}. While theoretical studies indicate that molecular imaging should be achievable using Bayesian algorithms such as the EMC algorithm \cite{Loh:2009} for near-perfect data simulated assuming currently available XFEL parameters \cite{Ayyer:2016}, this has yet to be demonstrated using experimental data containing realistic instrument background, sample variability and other experimental factors.  

This paper addresses the question of whether these above-mentioned experimental effects pose a fundamental roadblock to diffraction-pattern alignment and phasing algorithms in the low signal limit. We achieve this using experimental rather than simulated data. The approach taken is to start with experimentally measured data and progressively reduce the photon count to levels similar to those expected from smaller particles such as individual proteins. This process also mimics data that would be recorded from the same size particles using weaker X-ray pulses such as will soon be available with a high repetition rate from the LCLS-II upgrade.

We start from data collected by the SPI initiative from \SI{60}{\nm} PR772 viruses~\cite{Reddy:2017} to 8.5-nm resolution. Weak data was generated by keeping only a small, random fraction of photons from each experimental snapshot. These reduced data, or `diluted', patterns contain just a smattering of photons which often look like pure noise to the eye. In addition to diffraction from the virus particles, each diffraction pattern contains instrument background caused by a range of experimental sources. Any structure in the instrument background does not depend on particle orientation, thus after orientation determination this background appears as a spherically symmetric function incoherently added to the 3D Fourier intensities of the object. To account for this background, we develop a modified iterative phasing algorithm which isolates and retrieves this background while reconstructing the electron density, and also show that phase retrieval is robust to statistical noise.

The paper is set out as follows. The reconstruction pipeline and the results  of its application to the full data set are described in Section~\ref{sec:recon_steps}, and a set of metrics including the Fourier Shell Correlation (FSC) and Phase Retrieval Transfer Function (PRTF) for quantifying reconstruction resolution and fidelity are defined in Section~\ref{sec:metrics}. The experimental data sets are then subsampled by randomly selecting a fraction of photons in every frame, followed by orientation and phasing of the sparsified photon counts in Section \ref{sec:reduction}. The quality of the electron densities obtained using the subsampled data sets is evaluated and compared using the metrics of reconstruction quality defined in Section \ref{sec:metrics}. 

We find that the reconstruction quality persists for a significant reduction of data quantity: even when the signal is reduced by as much as 1/256, quality metrics show the virus electron density determined using \emph{ab initio} phasing is of almost the same quality as the high signal data.  This suggests that given sufficient number of single particle diffraction patterns from sub-10 nm biomolecules with current XFEL parameters (assuming a proportionate reduction in instrument background), or from 60-nm viruses with a pulse 256 times weaker, one can obtain reliable 3D electron densities with the methods presented here. In order to obtain higher resolution, many more patterns will be required to achieve sufficient statistics. This may soon be within reach with advancements in sample delivery methods as well as with high-repetition-rate XFEL sources such as the European XFEL and LCLS-II.


\section{\label{sec:experiment}Experiment description}
Diffraction snapshots of aerosolized PR772 viruses were collected at the Linac Coherent Light Source (LCLS) as described in \cite{Reddy:2017}. Briefly, diffraction patterns were recorded on a pnCCD detector in the AMO instrument at the LCLS~\cite{Ferguson:2015} at a photon energy of \SI{1.6}{\keV} with the detector placed \SI{586}{\mm} downstream from the X-ray-sample interaction point, giving a resolution of \SI{11.8}{\nm} at the center-edge of the detector and maximum resolution of \SI{8.4}{\nm} in the corner of the detector.   This data set is available for download from the Coherent X-ray Imaging database \cite{Maia:2012} as CXIDB 58.

The data set consists of \SI{14772} frames with an average signal level of \SI{395876} photons/frame.  For a \SI{60}{\nm} virus, the speckles were around 100 pixels wide. The pixels were therefore binned by a factor of 4 in both dimensions after photon conversion to reduce computational costs. Excluding bad pixels and the central speckle, where the detetor was often saturated, there were \SI{34783} photons/frame on average. There were on average 22.2 photons/speckle at the detector corner. 

Diffraction patterns were recorded at a repetition rate of 120 Hz, however only a small fraction of the X-ray pulses interacted with an object. These so-called ``hits'' included not only interactions with PR772 virus particles but also with water droplets, multi-particle clusters, and patterns with detector artifacts. Such spurious patterns need to be excluded from analysis. In \cite{Reddy:2017}, Reddy et al describes the classification of the single particle patterns using various machine learning methods, with the data for this study based on the classification by manifold embedding \cite{Yoon:2011} to obtain a data set consisting of \SI{14772} single virus diffraction patterns. 


\section{\label{sec:recon_steps}Reconstruction procedure}
\begin{figure}
    \centering
    \includegraphics[width=\columnwidth]{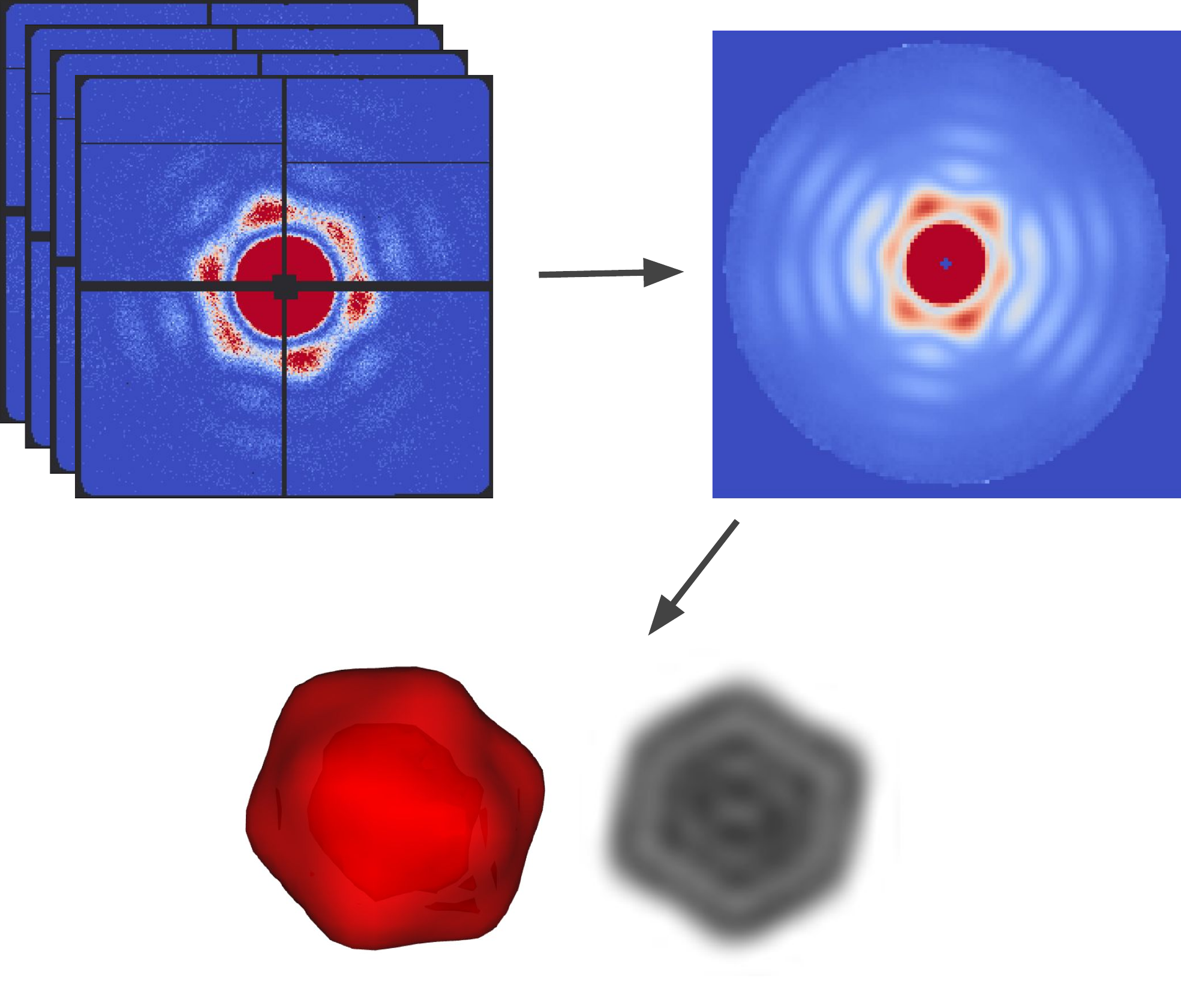}
    \caption{Reconstruction of the virus electron density from measured diffraction snapshots is a two step process.  First, the orientations of a set of noisy diffraction patterns of mostly identical objects in random orientations with variable incident fluence (top left) are determined to produce a 3D intensity volume (top right).  The three dimensional diffraction volume is then phased using a background-aware phase retrieval algorithm to arrive at the real space electron density (bottom). The electron density is shown as both an isosurface plot and a slice through the center of the object.}
    \label{fig:pipeline}
\end{figure}

The PR772 virus electron density was reconstructed in a two-step process, illustrated in Fig.~\ref{fig:pipeline} and detailed below. First, the orientations of a set of noisy diffraction patterns of mostly identical objects in random orientations with variable incident fluence were determined to produce a 3D intensity volume using the EMC algorithm~\cite{Loh:2009}. The three dimensional diffraction volume was then phased using a background-aware phase retrieval algorithm to arrive at the real-space electron density using a combination of the Difference Map~\cite{Elser:2003} and Error Reduction~\cite{Fienup:1978} algorithms.

\subsection{\label{sec:merging_steps}Alignment: Determining the 3D reciprocal space intensity distribution}
Orientation determination, alignment and scaling of the diffraction patterns into a 3D diffraction volume was performed using the \emph{Dragonfly} software~\cite{Ayyer:2016}.  Data was provided to \emph{Dragonfly} in photon counts since the pnCCD detector used in this experiment could resolve individual \SI{1.6}{\keV} photons. A Poisson noise model was therefore used in \emph{Dragonfly}. Both the orientation as well as a relative scale factor was estimated for each pattern to account for incident fluence fluctuations and variations in impact parameter of the virus relative to the beam. The predicted intensities on the detector for a given orientation were multiplied by this scale factor before calculating the probability distribution over orientations (PDOs). These scale factors were  updated every iteration using the current estimate for the PDO for each pattern. In order to avoid convergence issues due to the high signal per pattern, the PDO was raised to the power of the deterministic annealing parameter, $\beta$. This parameter was increased from 0.001 by a factor of $\sqrt{2}$ every 10 iterations. The detailed procedure used for this reconstruction is described in Appendix A.

\subsection{\label{sec:phasing_steps}Phasing: Iterative phase retrieval with background estimation}
The three dimensional diffraction volume from \emph{Dragonfly} was phased to arrive at the real space electron density using a background-aware iterative projection phase retrieval algorithm as described in Algorithm~\ref{alg:phase}. The update rule for this  algorithm consists of a modulus projection defined to incorporate a spherically symmetric background intensity which is incoherently added to the diffraction signal (``Background aware") in addition to a support constraint on the electron density consisting of a fixed number of voxels rather than a static mask (``Voxel number support"). 

The iterate $\Psi$ is comprised of both the real space density ${\rho(\mathbf{x})}$ and background ${B(\mathbf{q})}$
\begin{equation}
\Psi = \left\{\rho(\mathbf{x}), B(\mathbf{q})\right\}
\end{equation}
In practice this consists of two 3D volumes, one for the real-space electron density and the other for the square root of the background intensity. The calculated intensity is the sum of the intensity from the particle plus the background,
\begin{equation}
I_\text{calc}[\Psi](\mathbf{q}) = \left|\mathcal{F}[\rho](\mathbf{q})\right|^2 + B^2(\mathbf{q})
\label{eq:icalc}
\end{equation}
where $\mathcal{F}[\rho]$ is the discrete Fourier transform of the electron density $\rho$. The modulus projection rescales both terms by the ratio to the measured Fourier magnitude,
\begin{equation}
P_M[\Psi] = \left\{\mathcal{F}^{-1}\left[\sqrt{\frac{I_\text{meas}(\mathbf{q})}{I_\text{calc}(\mathbf{q})}} \mathcal{F}[\rho](\mathbf{q})\right], \sqrt{\frac{I_\text{meas}(\mathbf{q})}{I_\text{calc}(\mathbf{q})}}B(\mathbf{q})\right\}
\label{eq:pmod}
\end{equation}
where $I_\text{meas}(\mathbf{q})$ is the measured intensity. 

The support projection imposes two different constraints on the two halves of the iterate, $\rho$ and $B$. A constant $N$ is chosen at the beginning representing the number of voxels inside the particle for which the density is allowed to be non-zero. In this case we chose $N=2000$.  The modulus-squared electron density values are sorted and the highest $N$ are left unchanged while the rest are set to zero. The background intensities, $B(\mathbf{q})$, are replaced by the spherically symmetric version i.e. the intensities in each radial bin are replaced by their average. The derivation that both these operations are projections is given in Appendix B. Further details regarding masking and alignment of reconstructions from different random starting models are discussed in Appendix C.

\begin{algorithm}[H]
\begin{algorithmic}[1]
  \Function{ER}{$x$}
    \State \Return $P_M(P_S(x))$
  \EndFunction
  \item[]
  \Function{DM}{$x$}
    \State $\beta = 0.7$
    \State $f_M(x) = (1-1/\beta)P_M(x) + (1+1/\beta)x$
    \State $f_S(x) = (1+1/\beta)P_S(x) + (1-1/\beta)x$
    \State \Return $x + \beta\left[P_M(f_S(x)) - P_S(f_M(x)) \right]$
  \EndFunction
  \item[]
  \For{$i$ in 1 to 400}
    \State $\Psi_i \gets$ Uniform Random
    \State $\Psi_i \gets$ ER($\Psi_i$) (100 times)
    \State $\Psi_i \gets$ DM($\Psi_i$) (200 times)
    \State $\Psi_i \gets$ ER($\Psi_i$) (100 times)
  \EndFor
  \State Align all $\Psi_i$
  \State Calculate Phase Retrieval Transfer Function (PRTF)
  \State \Return Average over all aligned $\Psi_i$
\end{algorithmic}
\caption{Pseudo-code describing the iterative phasing of a 3D intensity volume with spherical background retrieval and a voxel number support. The modulus and support projections, $P_M$ and $P_S$ are described in Sec.~\ref{sec:recon_steps}.}
\label{alg:phase}
\end{algorithm}

\subsection{\label{sec:phased_all}Reconstruction from the full data set}
The results of applying the above two-step reconstruction method to all \SI{14772} patterns are shown in Fig.~\ref{fig:pipeline}. The 3D intensity shows strong icosahedral symmetry even though this constraint was not enforced during the reconstruction. The resolution corresponding to the edge of the spherical volume of intensities is \SI{8.4}{\nm}. After iterative phasing, the electron density shown in the bottom row was obtained. The contour plot shows an icosahedron with bulges at each vertex while a slice through the object centre shows the presence of a double-walled shell with a slight reduction in density just inside the outer shell, consistent with other treatments of the data~\cite{Kurta:2017,Rose:2018}.


\section{\label{sec:metrics}Quantifying reconstruction quality}
A set of quantitative metrics are required in order to compare reconstructions and assess overall reconstruction quality, for reconstructions of both the full and diluted data sets. We used two metrics established in the literature, which we define in this section for clarity, and applied them to the reconstruction performed with the full data set described above.

\subsection{\label{sec:fsc}``Gold-standard'' cross correlations}
The first of these metrics, inspired by cryo-electron microscopy, involves a slight change in the analysis pipeline itself. The `gold-standard` Fourier shell correlation from CryoEM~\cite{Henderson:2012} calls for the separation of the dataset into two equal halves. Each half is analyzed independently, the final volumes rotationally aligned, and the relative agreement is calculated as a function of resolution using the Fourier Shell Correlation (FSC) metric:
\begin{equation}
\mathrm{FSC}(q) = \operatorname{Re}\left[\frac{\sum\limits_{|\mathbf{q}_i| = q} F_1(\mathbf{q}_i) F_2^*(\mathbf{q}_i)}
{\sqrt{\sum\limits_{|\mathbf{q}_i| = q} |F_1(\mathbf{q_i})|^2}\sqrt{\sum\limits_{|\mathbf{q}_i| = q} |F_2(\mathbf{q_i})|^2}}\right]
\end{equation}
where $F(\mathbf{q}) = \mathcal{F}[\rho](\mathbf{q})$. In practice, the FSC is calculated in $q$ bins which are shells of a certain thickness.

A similar correlation can also be calculated between the two half-dataset intensities. In order to increase the sensitivity of the correlation, the mean is subtracted in each resolution shell before calculating the cross-correlation i.e. a Pearson correlation coefficient is calculated in each shell independently.
\begin{equation}
\mathrm{CC}_{1/2}(q) = \frac%
  {\sum\limits_{|\mathbf{q}_i| = q} \left(\mathrm{I}_1 - \overline{\mathrm{I}_1}\right) \left(\mathrm{I}_2 - \overline{\mathrm{I}_2}\right)}%
  {\sqrt{\sum\limits_{|\mathbf{q}_i| = q} \left(\mathrm{I}_1 - \overline{\mathrm{I}_1}\right)^2}%
   \sqrt{\sum\limits_{|\mathbf{q}_i| = q} \left(\mathrm{I}_2 - \overline{\mathrm{I}_2}\right)^2}}
\end{equation}
where $\mathrm{I}_k$ is shorthand for $\mathrm{I}_k(\mathbf{q}_i)$ and $\overline{\mathrm{I}_k}$ is the mean intensity in the resolution shell $\overline{\mathrm{I}_k(q)}$. The increased sensitivity due to subtracting the mean is most apparent when there is spherically symmetric background in the intensity reconstruction, as is the case here.

\subsection{\label{sec:prtf}Phase retrieval transfer function (PRTF)}
The other metric is the phase retrieval transfer function (PRTF)~\cite{Shapiro:2005}. This metric measures the reliability of iterative phasing by (in effect) averaging complex values over may instances of the phasing process. 

The first step in the calculation of this metric is to reconstruct a large number of independent density volumes from different random starting guesses. At any given reciprocal-space voxel, $\mathbf{q}$, the argument of the complex Fourier transform of the density (the phase) can be slightly different in each random start. The value of the PRTF at that voxel is the complex sum of the unit complex numbers whose argument is the phase, $\phi$:
\begin{equation}
\mathrm{PRTF}(\mathbf{q}) = \frac{1}{N}\left|\sum_{n=1}^N e^{i\phi_n}\right|
\end{equation}
where there are $N$ independent density volumes. By convention, the azimuthal average of the PRTF is reported as a function of the radial coordinate $\left|\mathbf{q}\right|$. As described in Sec.~\ref{sec:phasing_steps}, the different reconstructions must be aligned in real-space before calculating the average. A shift in real space is equivalent to a phase ramp which will significantly lower the PRTF. An uncorrected central inversion will negate the phase, leading to a similar reduction~\cite{Marchesini:2006}.

One weakness of the PRTF is that it can be unjustifiably high if the support volume is chosen to be too small. As an extreme case, if the support consists of only one voxel, the PRTF (after alignment) will be unity everywhere even though the reconstruction is very poor. One should therefore have a slightly larger support mask which includes some voxels with low density. In the reconstructions performed here, the support volume (2000 voxels) is significantly larger than the nominal volume of a regular icosahedron with a size corresponding to the fringe spacing (which would be 1497 voxels).

We calculate the PRTF from 400 independent reconstructions. This number is important because it needs to be large enough for the PRTF to converge and the voxels with irreproducible phases to average down. Consider for example the case where the phases are completely random, in which case the sum is a 2D random walk in the complex plane with a fixed step size which has an average distance from the origin of $\sqrt{N}$ after $N$ steps. Thus, the expected lower bound on the PRTF if $N$ reconstructions are averaged is $1/\sqrt{N}$, which is 0.05 for the case of 400 the case here. In keeping with convention, the threshold value to determine the reproducible resolution is considered to be $1/e = 0.37$.


\begin{figure}
\centering
\includegraphics[width=\columnwidth]{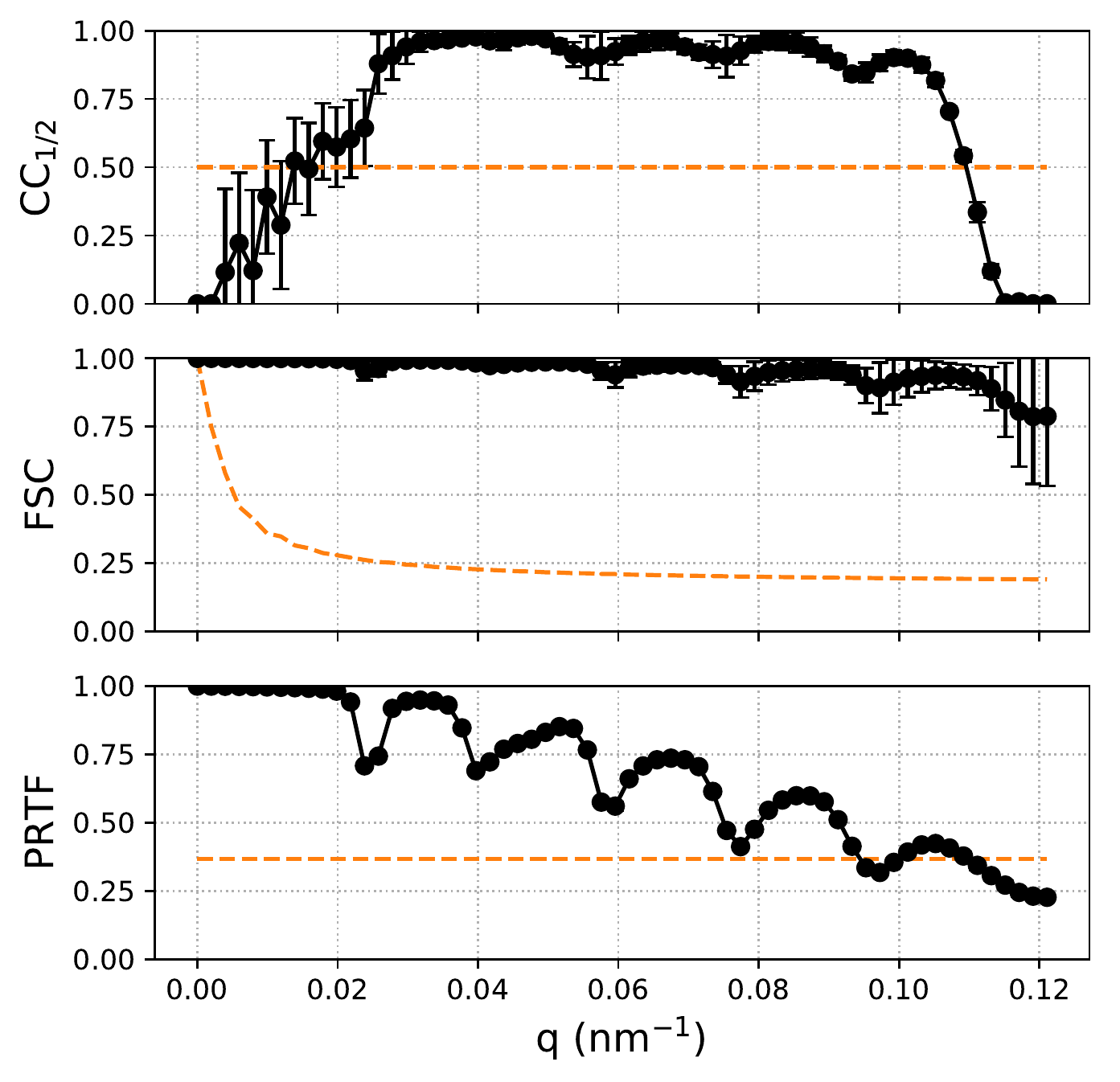}

\caption{Reconstruction metrics for the full data set as a function of $q$. Top: Fourier Shell Correlation (FSC) plot with the dashed line showing the half-bit threshold. Middle: Intensity CC$_{1/2}$ plot with the dashed line showing the 0.5 cutoff. Bottom: Phase Retrieval Transfer Function (PRTF) plot with the customary $1/e$ cutoff. Error bars represent the standard deviation across 10 random starts.}
\label{fig:full-metrics}
\end{figure}

\subsection{\label{sec:full_recon}Metrics applied to full data reconstruction}
We applied the metrics defined above to the reconstructed intensity and electron density calculated using the procedure described in Sec.~\ref{sec:recon_steps}.  For the FSC and CC$_{1/2}$ calculations, frames were split into and odd and even halves containing the 1st, 3rd, 5th... and 2nd, 4th, 6th... patterns respectively. This procedure of splitting is chosen in order for both halves to be similarly affected by slowly varying drifts in the experiment. It is also sufficiently random because the ``hits" themselves are a random subset of all the patterns collected. 

The FSC and CC$_{1/2}$ plots are shown in Fig.~\ref{fig:full-metrics}. The crystallographic definition of $q$ is used with the full-period resolution, $d = 1/q$. 
Each of the metrics gives a slightly different estimate of the resolution of the reconstruction. from the half-bit FSC criterion standard common in cryo-electron microscopy~\cite{VanHeel:2005}, the resolution is \SI{8.75}{\nm}, while using the CC$_{1/2} = 0.5$ cutoff, the intensities are reproducibly reconstructed to a resolution of \SI{9.02}{\nm}. The purely phasing metric, PRTF, suggests that the resolution is \SI{10.9}{\nm} for both the even and odd data sets.  The oscillations apparent in the PRTF plot, which manifest from fringe intensities in the data, further reveal how resolution determined by the PRTF metric can be dramatically affected by whether or not values in one of the local minima happen to lie above or below the 0.37 threshold value. That the resolution estimates differ is not surprising given that different quantities are being measured, and suggests that one should be cautious when reporting a single resolution number.  The difference between values further suggests being very conservative with the precision to which resolution is quoted in publication:  the mean resolution estimated above is \SI{9.5}{\nm} with a standard deviation of \SI{1.1}{\nm}, in which case quoting resolution to three significant figures is certainly not appropriate.  One should further be careful comparing resolution between publications to make sure that the same values are being compared.

\section{\label{sec:reduction}Results}
We now turn our attention to the effect of reducing the amount of data on reconstruction quality using the analysis pipeline described in Section~\ref{sec:recon_steps}.  Data quantity is reduced in one of two ways. Diffraction patterns can be made weaker to simulate the effect of imaging smaller particles or the effect of a lower intensity X-ray beam. This has two effects: firstly orientation determination is expected to become harder as there is less information in each pattern from which to determine the orientation, and secondly the signal-to-noise ratio of the reconstructed 3D intensities is reduced making phase retrieval more challenging.  Alternatively, the number of diffraction patterns can be reduced to simulate the effect or a smaller data set consisting of fewer diffraction patterns of the same signal strength.  Computationally reducing the data in this way avoids confounding factors from working with different data sets collected at different times under potentially different experimental conditions.


\subsection{\label{sec:red_patterns}Reducing diffraction pattern intensity}
\begin{figure}
\centering
\begin{tabular}{ c c }
  \includegraphics[width=0.5\columnwidth]{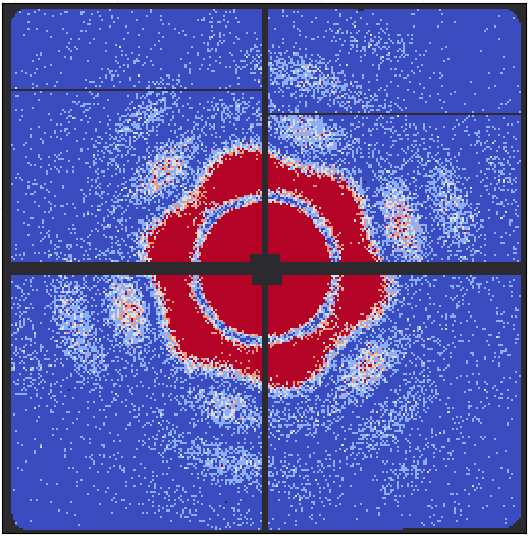} & \includegraphics[width=0.5\columnwidth]{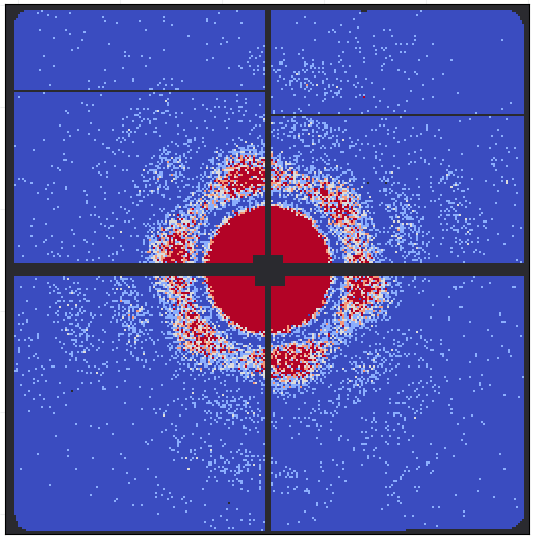}\\
  \small (a)  & \small (b) \\
  \includegraphics[width=0.5\columnwidth]{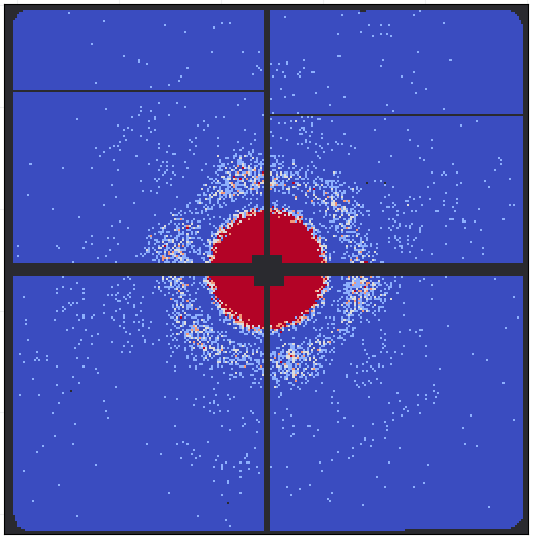} & \includegraphics[width=0.5\columnwidth]{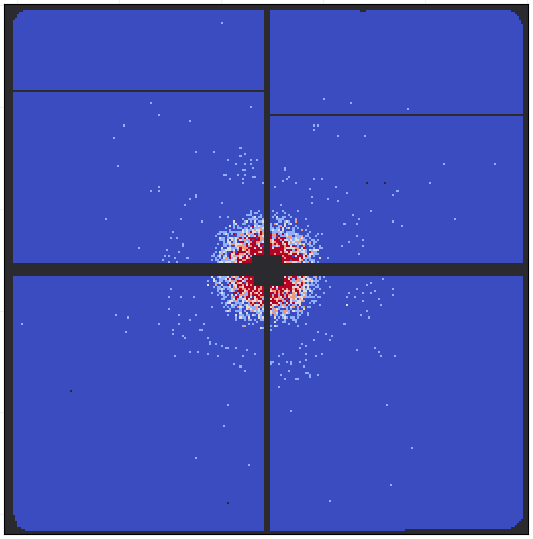}\\
  \small (c)  & \small (d) \\
\end{tabular}
\caption{Four versions of the same diffraction pattern showing the reduction of photons/pattern by a given selection probability, $p$. In each case, the color scale maximizes at 4 photons per pixel. (a) Original pattern (b) $p=1/4$ (c) $p=1/16$ (d) $p=1/256$}
\label{fig:dilution}
\end{figure}

To simulate measurement of weaker diffraction patterns we computationally reduced the number of photons in each image to produce diffraction patterns with fewer photons drawn from the same experimental data sets.
Reducing the number of photons in each diffraction pattern was done by applying a Bernoulli process to each photon with a certain probability to keep or discard the photon. These selection fractions, $p$, were reduced from $2^{-1}$ to $2^{-10}$ in steps of powers of two. Due to the Poisson nature of the photon counting statistics, this simulates the effect of a factor $p$ weaker incident pulse. The effect of applying this process to a particular diffraction pattern is shown in Figure~\ref{fig:dilution}.  The average number of photons per frame after photon dilution is shown in Table \ref{table:dilution}, from which it can be seen that photon counts per frame decreases from nearly 35,000 photons per frame at full strength to only 33 photons per frame when diluted to 1/1024 strength.

Reconstruction of the 3D intensity from weakened data was performed in the same manner as previously described for all data sets using identical \emph{Dragonfly} reconstruction parameters for all data sets except for the schedule of the deterministic annealing parameter $\beta$. A low value of $\beta$ was not necessary when the signal level was low since this parameter acts to solve convergence issues for very high signals by broadening the PDOs. Appendix A contains details of the parameters for each subset. The 3D intensities from \emph{Dragonfly} were phased with identical parameters in every case to generate electron densities.  Each reduced data set was split into two halves and independently reconstructed in order to calculate the ``gold-standard" FSC and CC$_{1/2}$, and this whole process was repeated 10 times to obtain error bars on the metrics.  

\begin{figure}
\centering
\begin{tabular}{ c c }
\includegraphics[width=0.45\columnwidth]{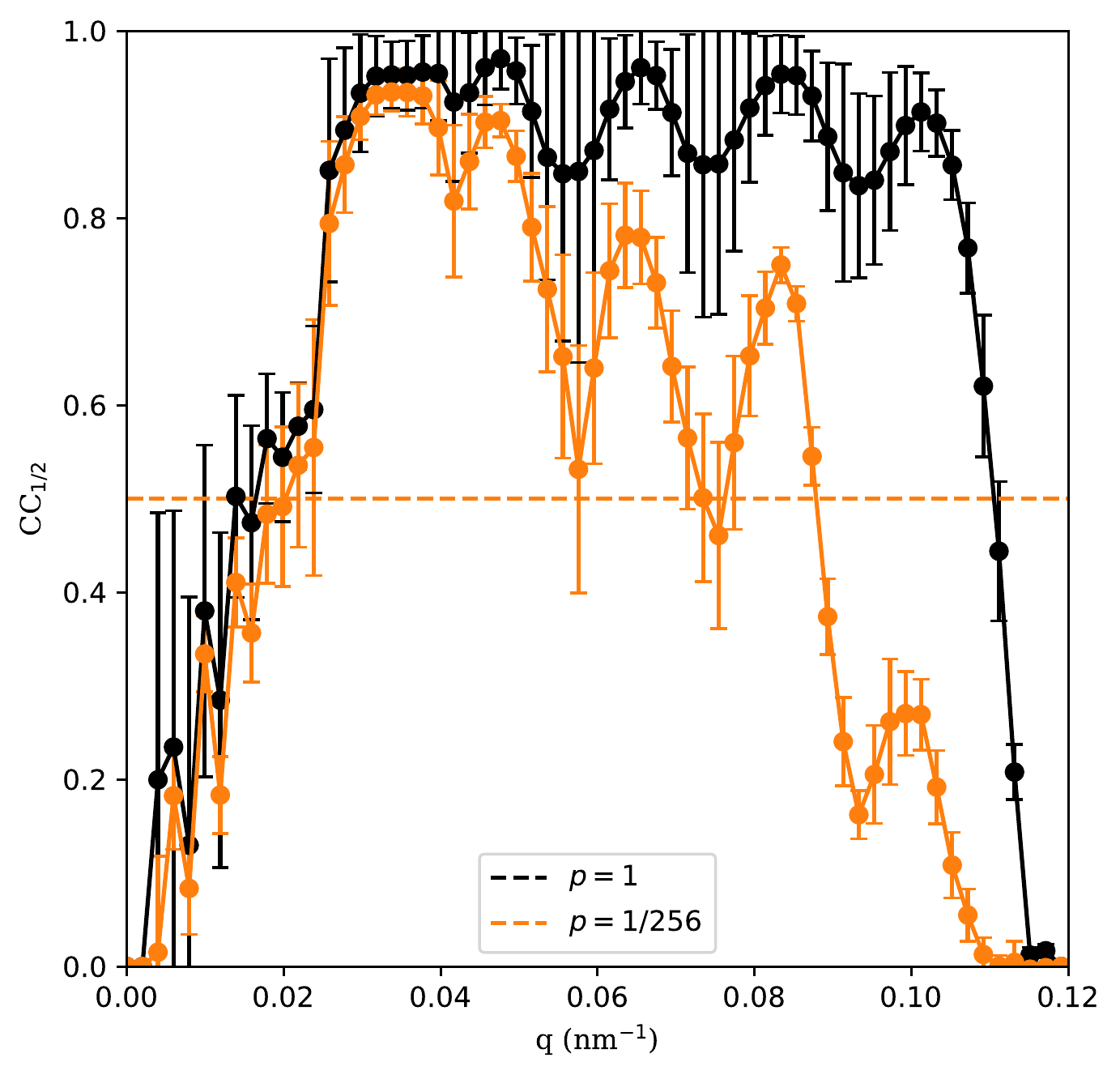} & \includegraphics[width=0.5\columnwidth]{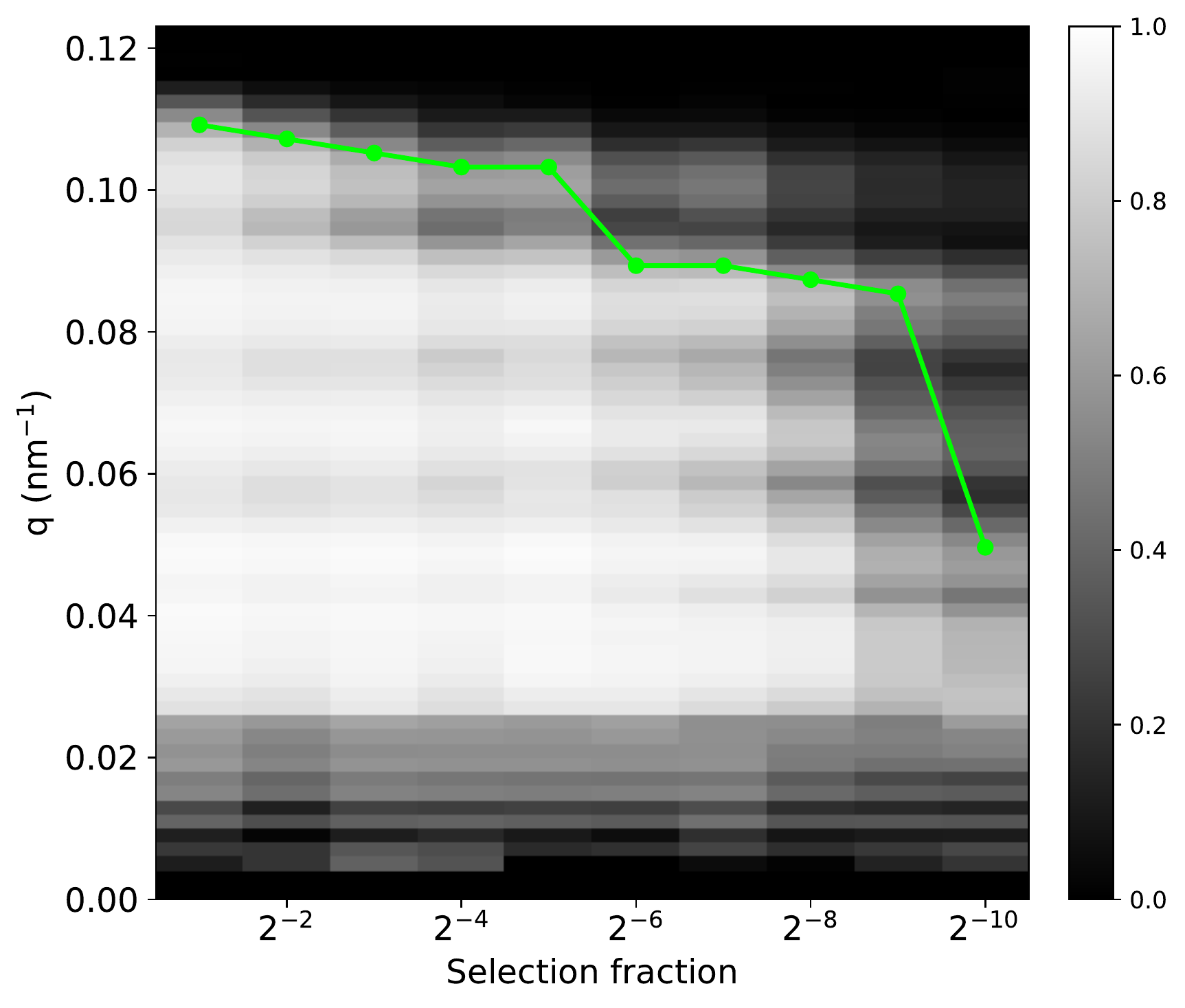} \\
\small (a) & \small (b) \\
\includegraphics[width=0.5\columnwidth]{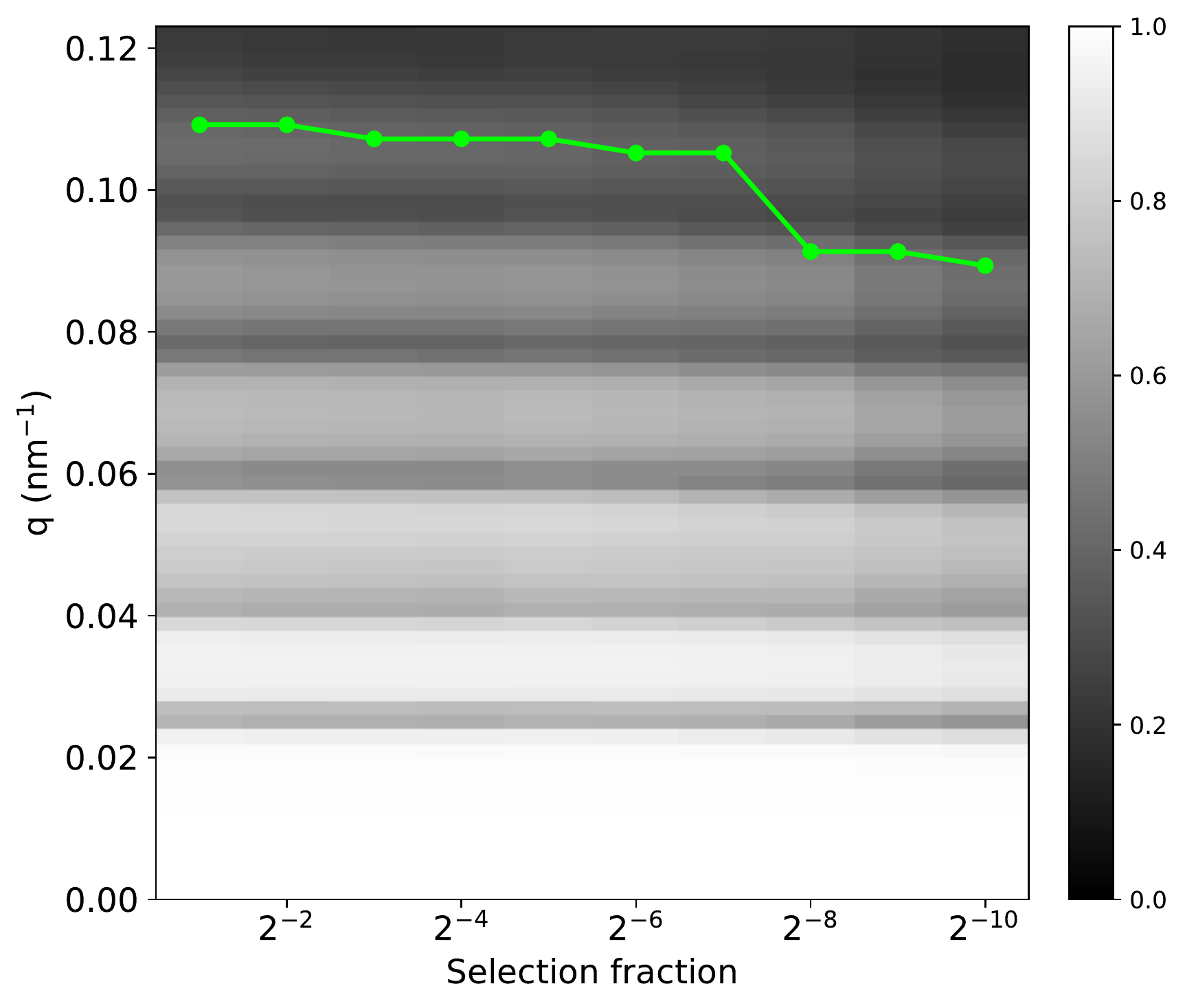} & \includegraphics[width=0.5\columnwidth]{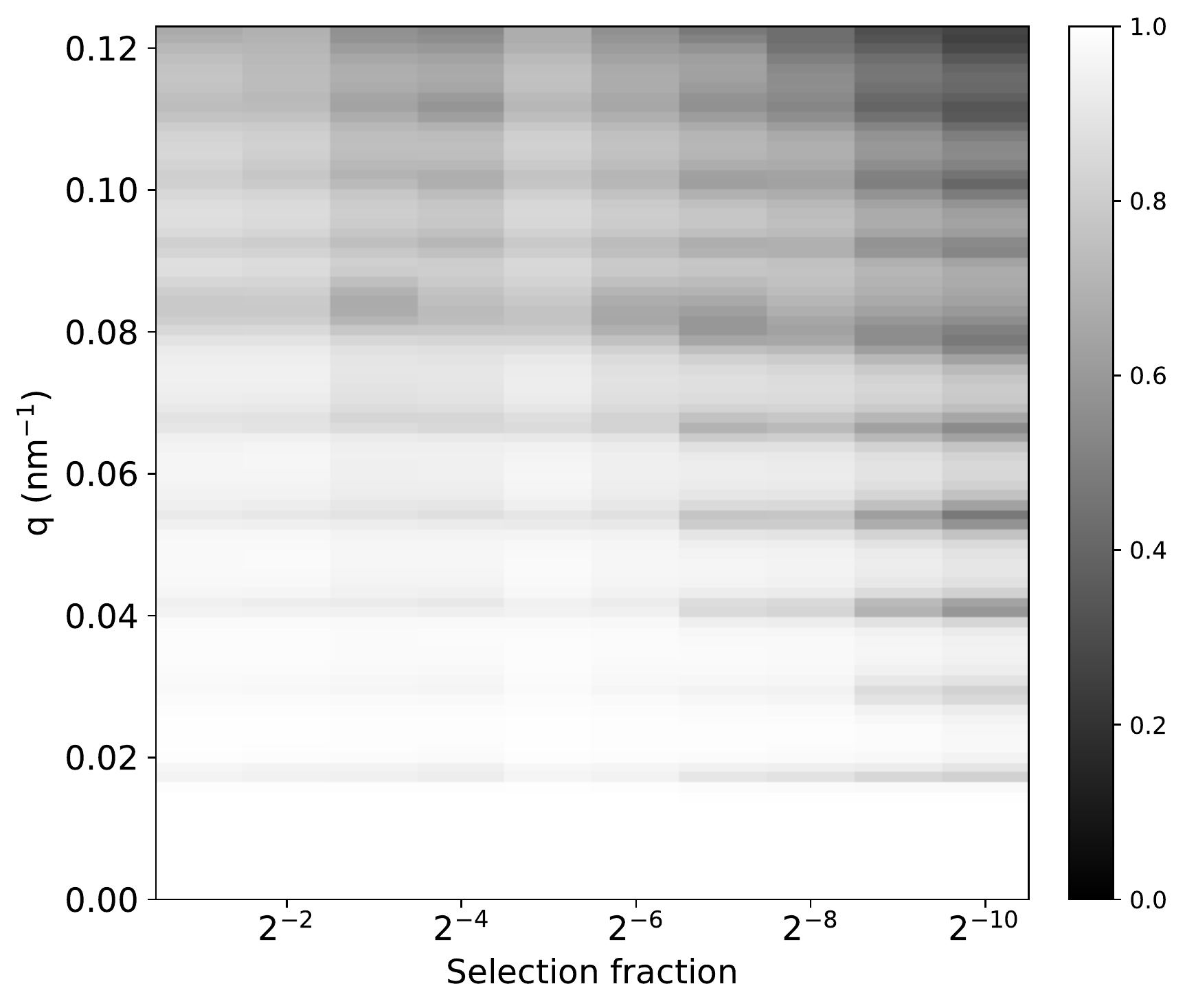}  \\
\small (c) & \small (d)
\end{tabular}
\caption{Dependency of reconstruction metrics on selection fraction. (a) Plots of CC$_{1/2}$ vs $q$ for the full data set and for a selection fraction of $p=2^{-8}=1/256$. Error bars represent the standard deviation across 10 different random half datasets and the dashed line represents the CC$_{1/2}=0.5$ cutoff.  To represent dependence on selection fraction, we plot the metric in grayscale versus both selection fraction and $q$ in panels (b)-(d) with the color representing the metric value.  (b) CC$_{1/2}$, the green dashed line shows the $q$ for the CC$_{1/2} = 0.5$ cutoff; (c) PRTF, dashed line shows the typical PRTF$=1/e$ cutoff, and (d) FSC, where the metric never went below the standard half-bit criterion. Each plot is the average of 10 random subsets.}
\label{fig:split_metrics}
\end{figure}

The results of reducing signal strength are summarized in Fig.~\ref{fig:split_metrics}. In Fig.~\ref{fig:split_metrics}(a) we plot one metric, CC$_{1/2}$, as a function of $q$ for both the full data set and a selection fraction of $p=2^{-8}=1/256$.  Fig.~\ref{fig:split_metrics}(a) shows that the reconstruction from the reduced data shows a slightly decreased quality metric compared to the full data set.

In order to summarise the results as a function of resolution for many different photon dilution levels, in Figs.~\ref{fig:split_metrics}(b)--\ref{fig:split_metrics}(d) we plot each metric in grayscale versus both selection fraction and $q$, where color represents the metric value. The green dashed line in Fig.~\ref{fig:split_metrics}(b) marks the somewhat arbitrarily chosen CC$_{1/2} = 0.5$ cutoff, and shows how the resolution of the intensity reconstruction becomes progressively worse as $p$ is reduced. One cause of this reduction is just the graininess of the reconstruction due to insufficient total signal. Similarly the green line in Fig.~\ref{fig:split_metrics}(c) represents the the typical PRTF$=1/e$ cutoff.  The step decrease in resolution shown by the PRTF in  Fig.~\ref{fig:split_metrics}(c) occurs when the overall PRTF decreases to the point where the next local minima falls below cutoff threshold, Fig.~\ref{fig:full-metrics}.  The resolution estimated by each metric is tabulated in Table \ref{table:dilution}.

From the metrics alone one immediately notices that the electron densities do not suffer from such a drastic falloff in resolution at very low signal. In effect, the support constraint during phasing restores the smoothness of the speckles even when  the total number of photons per 3D speckle (Shannon voxel) is low, partially negating the effect of insufficient total signal. For the highest photon dilution ($p=1/1024$), the average signal level used to determine the orientations is just 33.9~photons/frame.

\begin{table}
\centering
\caption{Data statistics as a function of selection fraction. The photons per frame described in the second column refers to photons outside the central speckle. The last three columns give the resolution in nanometers according to the standard cutoff criteria for the respective metric.}
\begin{tabular}{l r c c c c}
     \hline
     Fraction & ph/fr & Frames & CC$_{1/2}$ & PRTF & FSC \\
     \hline \\
     $1$ & \SI{34783.2} & \SI{14772} & 9.02 & 10.19 & 8.75 \\
     $1/2$ & \SI{17349.3} & \SI{14772} & 9.16 & 9.16 & 8.75 \\
     $1/4$ & \SI{8674.5} & \SI{14772} & 9.33 & 9.16 & 8.75 \\
     $1/8$ & \SI{4337.3} & \SI{14772} & 9.50 & 9.33 & 8.75 \\
     $1/16$ & \SI{2168.6} & \SI{14772} & 9.69 & 9.33 & 8.75 \\
     $1/32$ & \SI{1084.3} & \SI{14772} & 9.69 & 9.33 & 8.75 \\
     $1/64$ & \SI{542.2} & \SI{14772} & 11.2 & 9.50 & 8.75 \\
     $1/128$ & \SI{271.0} & \SI{14772} & 11.2 & 9.50 & 8.75 \\
     $1/256$ & \SI{135.5} & \SI{14772} & 11.4 & 10.9 & 8.75 \\
     $1/512$ & \SI{67.8} & \SI{14772} & 11.7 & 10.9 & 8.75 \\
     $1/1024$ & \SI{33.9} & \SI{14772} & 20.1 & 11.2 & 8.75 \\
\end{tabular}
\label{table:dilution}
\end{table}

\begin{figure}
\centering
\begin{tabular}{c}
    \includegraphics[width=3in]{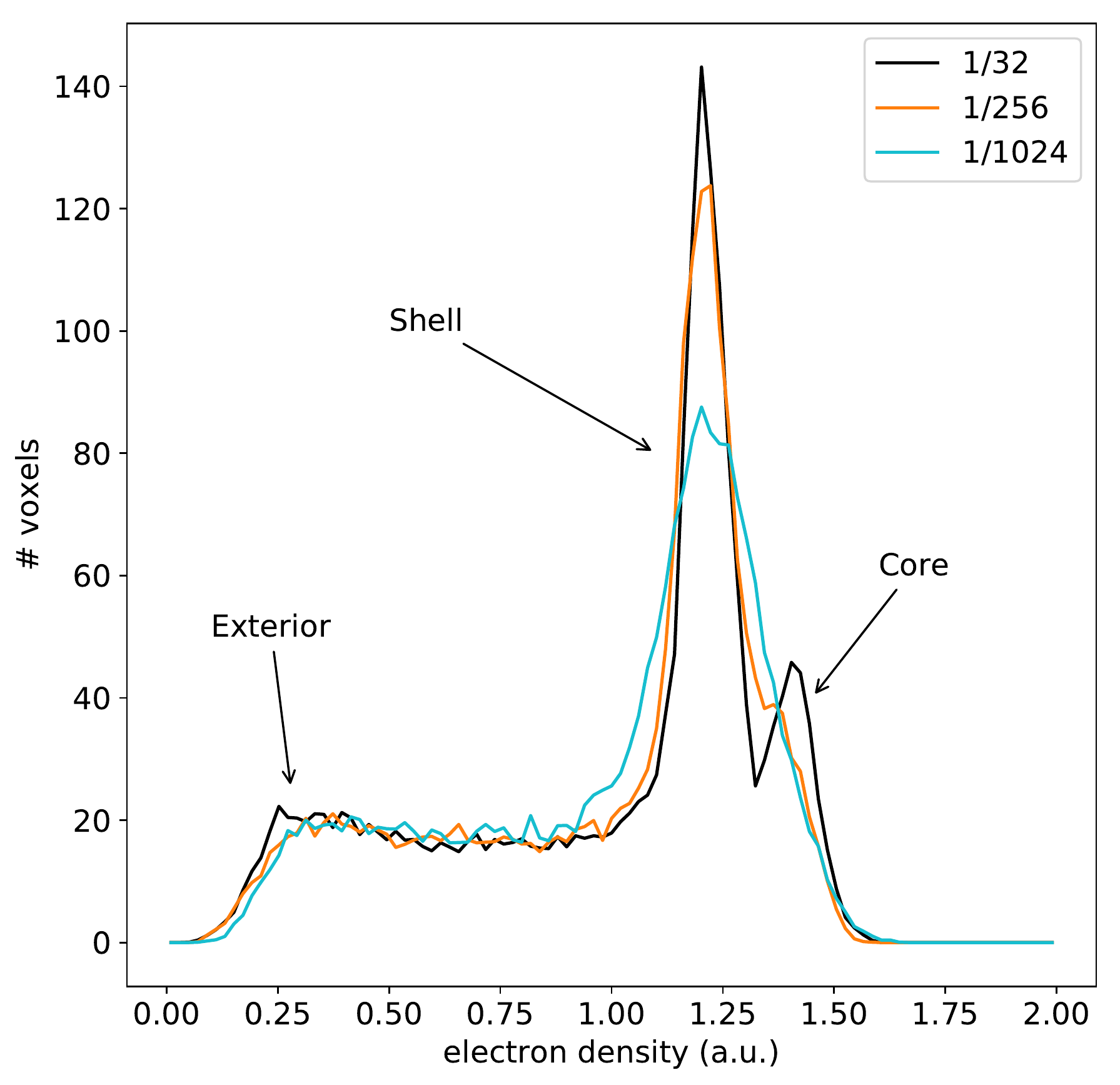} \\
    (a) \\
    \includegraphics[width=3in]{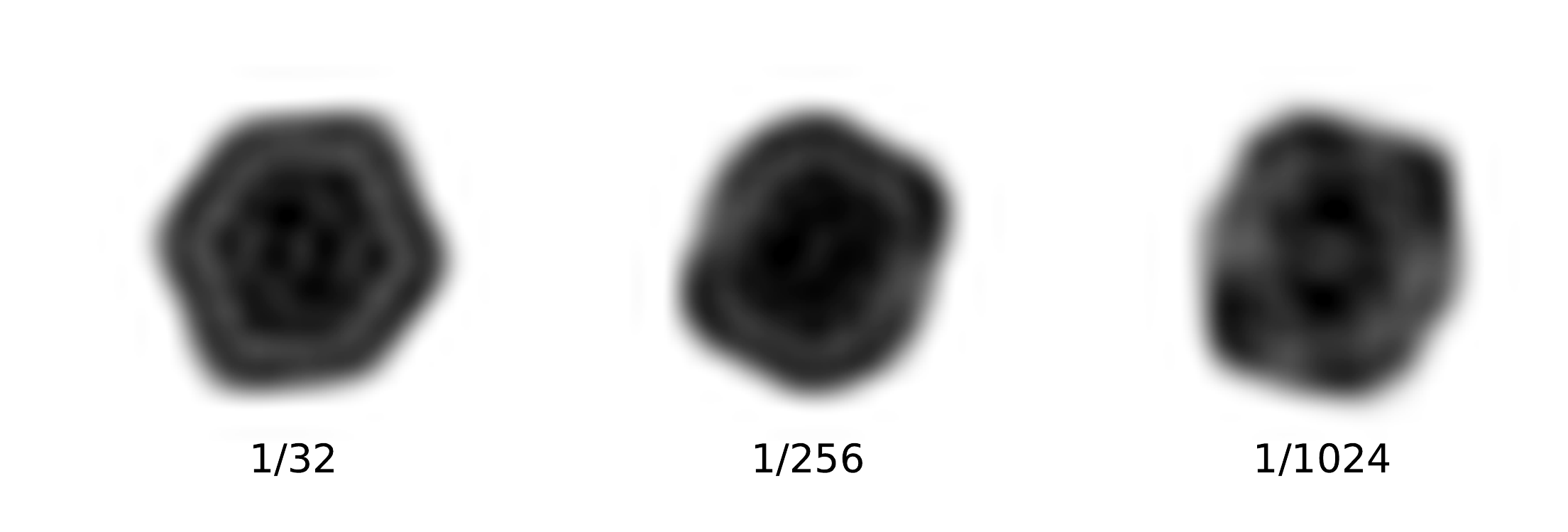} \\
    (b) \\
\end{tabular}
\caption{(a) Histogram of reconstructed electron densities for three different selection fractions. The voxels with low densities are present because the support is slightly larger than the particle. At higher photon counts, one can see a separation between the higher densities in the core of the virus compared to the capsid shell. This distinction disappears at the very low signal levels corresponding to $p=1/1024$. (b) Slices through representative electron densities with the same selection fractions. One can see the gradual disappearance of the double-shell structure with reducing fraction.}
\label{fig:dens_hist}
\end{figure}

We also studied the effect of reducing data on the histogram of electron density values retrieved in real space. Figure~\ref{fig:dens_hist} shows the histogram of electron densities inside the support mask for three different selection fractions. The plots are averaged over the 20 phasing runs for each fraction (10 random subsets and two halves per subset). The histograms clearly show the degradation in quality as signals are reduced, with the average reconstructed particle tending towards a uniform icosahedral blob with no internal structure. Additionally, the presence of the low density voxels is reassurance that the support was not too tight and the calculated PRTF not artificially high. For selection fractions above $1/32$, the histograms and densities were nearly identical, and are hence not shown for clarity. The difference in electron density histograms suggests that differences in the real space electron density may not be entirely reflected in all of the reconstruction metrics, and that metric cutoff values used to assess resolution may on their own paint a partial picture of reconstruction quality.  

\subsection{\label{sec:red_num_patterns}Reducing number of patterns}
An alternative method of reducing the total number of measured photons is be to select a random subset of full intensity diffraction patterns. By this method one approaches the limit of a few bright patterns.   

\begin{figure}
    \centering
    \includegraphics[width=0.9\columnwidth]{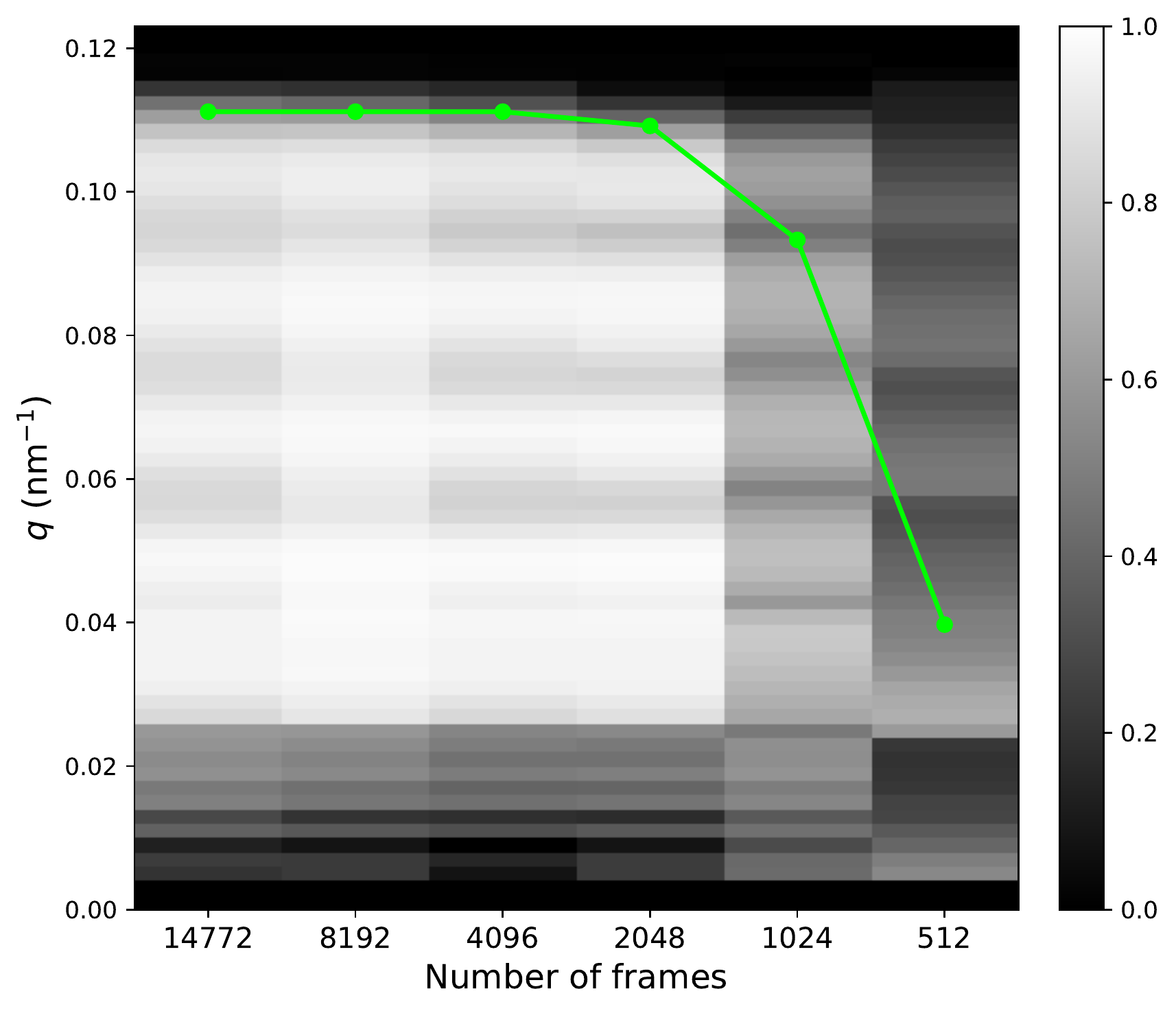}
    \caption{Intensity CC$_{1/2}$ vs $q$ plots as a function of number of frames in the data set. Like in Fig.~\ref{fig:split_metrics}, each column represents a plot or a different number of frames.}
    \label{fig:cc_nframes}
\end{figure}

From the total number of \SI{14772}, 10 random subsets were generated with 8192, 4096, 2048, 1024 and 512 patterns respectively. Each of these subsets was split into two halves (the even and odd patterns) and independently reconstructed. The CC$_{1/2}$ plots for the intensity reconstructions for each of the subsets is shown in Fig.~\ref{fig:cc_nframes}. Using this approach the metrics remain largely unaffected provided more than 2048 patterns in total are used (1024 in each half data set), indicating that the reconstruction was very stable and supports the hypothesis that there was more than enough data for this resolution. However, with 1024 frames (512 frames in each half), the reconstruction failed 4 out of the 20 times. What happens in this case is that if the number of patterns is reduced too much, they do not fill the 3D reciprocal space volume, leading to artifacts in orientation determination. Since a unique assignment of orientation for just 512 patterns would be insufficient to fully populate reciprocal space, the reconstruction only succeeds due to the PDOs being broad when $\beta$ is low. Even so, there are times when the 3D intensity collapses into a single, or a few planes: orientation determination effectively fails and all frames are assigned to one or a few orientations. Fortunately, this failure mode is easy to identify and exclude from averaging. The failed reconstructions have been retained in this work for the sake of completeness. Other algorithms which use additional constraints on the intensity, from a restricted real-space support, or from additional point-group symmetries, may have better performance in this limit of a few very bright patterns.

\section{\label{sec:discussion}Discussion}
By sub-sampling the experimental data from PR772 viruses measured in \cite{Reddy:2017}, we show that the reconstruction quality is essentially same as from the full data set with as few as 135 relevant photons/pattern, corresponding to 0.087 photons/speckle at the detector corner. This approaches the limits of prior work using simulated data \cite{Neutze:2000,Loh:2009,Ayyer:2016} or proof-of-principle experiments under highly controlled conditions not realistic for single particle imaging \cite{Philipp:2012,AyyerP:2015}.  By way of contrast, the results here are based on data derived from experimental measurements on PR772 viruses incorporating particle variability and instrument background, demonstrating that the signal required for X-ray single particle imaging under realistic conditions is much lower than previously demonstrated especially in terms of the number of scattered photons required per frame.  

From this numerical experiment we conclude that current SPI algorithms should be  capable of processing experimental single particle diffraction patterns when the photon flux in the X-ray focus is 256 times smaller than currently available at LCLS for particles of the same size as PR772.  Furthermore, algorithms appear to be more robust for the case of many weak hits than a small number of very strong hits. The extension of this method to smaller particles is not so direct. In order for this analysis to also hold for the case where the particle volume is reduced by the same factor, one requires that the parasitic scatter is also proportionately reduced. At higher photon energies, significantly lower background has already been achieved ~\cite{Munke:2016} than present in this data set. Thus, one strategy for the future direction of the field may be to move to hard X-ray instruments where one has reduced scattering cross section (factor 20 lower for 7 keV vs 1.6 keV, as was the case here) but possibly much lower background. 

From this analysis we also conclude that analysis algorithms on their own are not the current limiting factor for SPI imaging.  Low background data collection has already been demonstrated in the data set of \cite{Munke:2016} to 6\AA{} resolution. Unfortunately there were insufficient hits from the entire beamtime for a reconstruction to be feasible. The work here suggests that signal levels may have been adequate had sufficient single-particle diffraction patterns been collected.  This points to the need to further develop methods for introducing single particles into the X-ray focus in sufficient density to make sufficient measurements at high resolution. Indeed, this could currently be one of the main factors limiting further progress in SPI imaging. Another key conclusion is that further work is needed in the area of single particle diffraction pattern classification to achieve similar noise tolerance as orientation determination, for which the efficacy of machine learning techniques in the limit of low signal still needs to be explored. This result bodes well for the prospects of single particle flash X-ray imaging to near-atomic resolution at high repetition rate XFELs like the European XFEL and LCLS-II and may help guide future XFEL and instrument design. 

\begin{figure}
\centering
\begin{tabular}{c @{\qquad} c }
  \includegraphics[width=0.5\columnwidth]{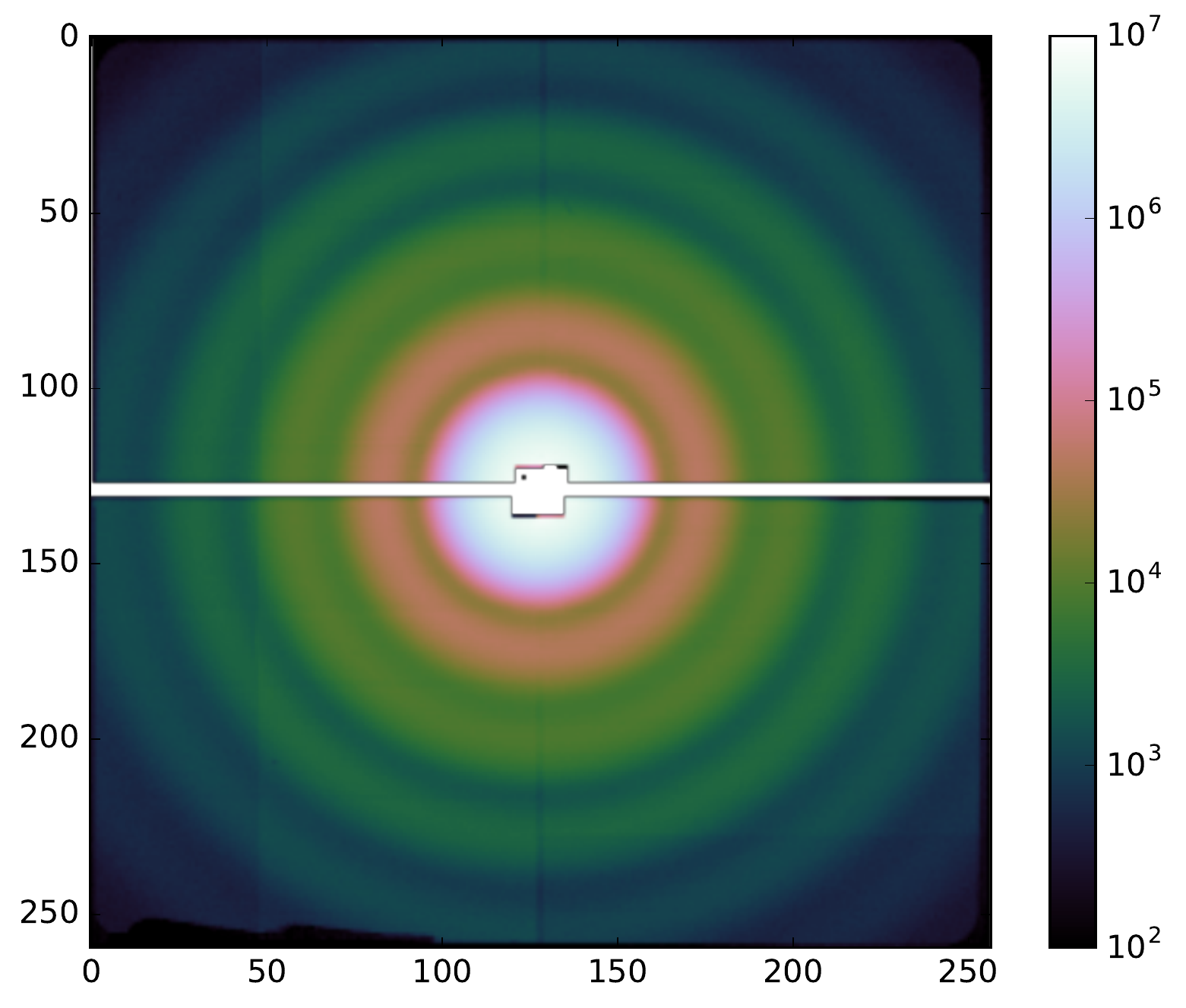} &
  \includegraphics[width=0.43\columnwidth]{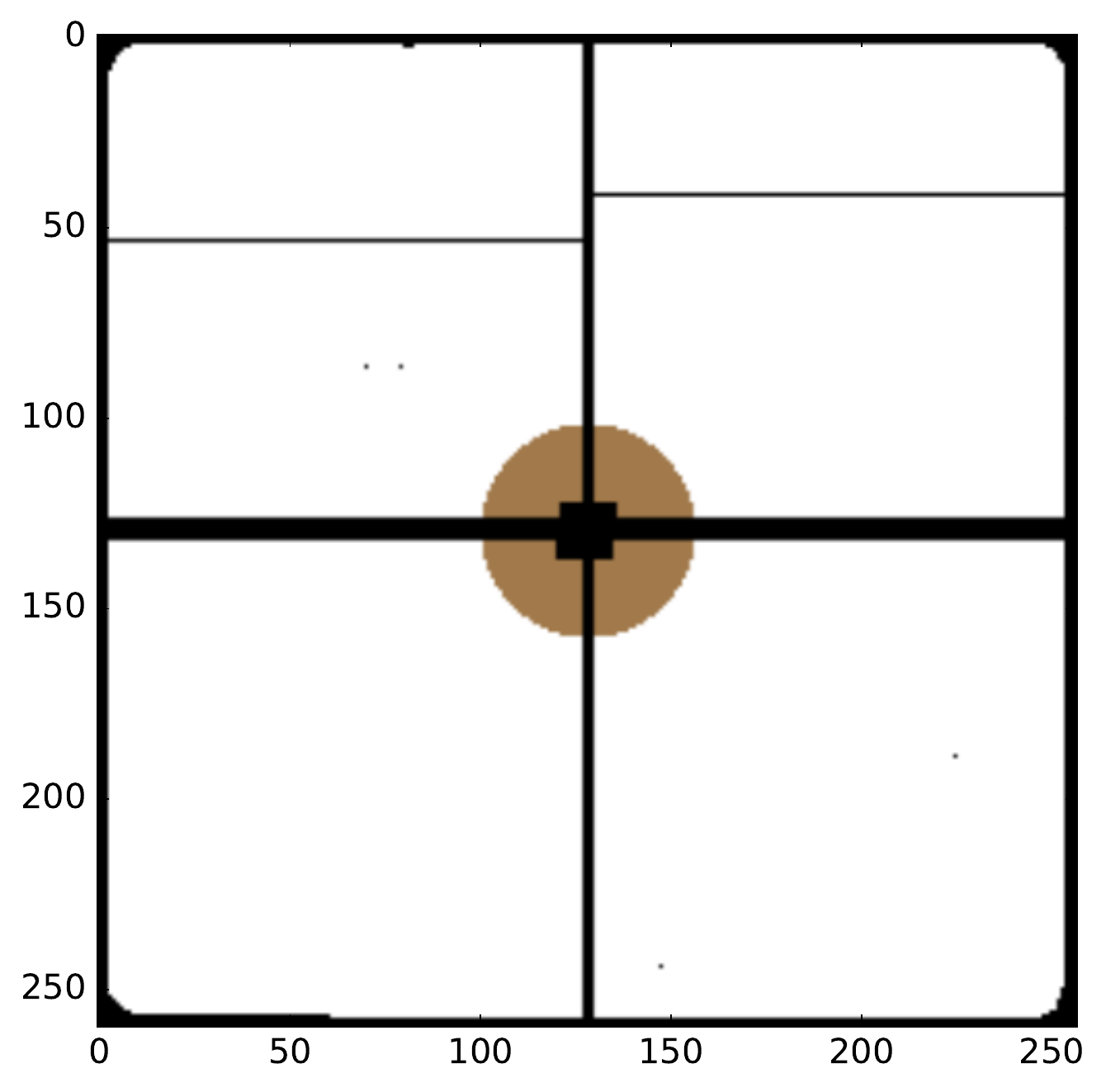} \\
  \small (a) & \small (b)
\end{tabular}
\caption{(a) Virtual powder sum of all \SI{14772} patterns, shown with a logarithmic color scale. (b) Detector mask used in orientation determination and intensity reconstruction. The `black' pixels were ignored completely. The `ochre' pixels were used to calculate the average intensity in 3D but not to calculate the orientations. The `white' pixels were used for both orientation and average intensity calculations.}
\label{fig:powder_mask}
\end{figure}

\section*{\label{sec:app_dragonfly}Appendix A: Intensity reconstruction details}
This appendix gives the detailed steps applied to reconstruct the intensity volume from the full dataset with \SI{14772} frames shown in Sec.~\ref{sec:full_recon}. A similar procedure was used for the reduced data set reconstructions whose results are described in Sec.~\ref{sec:reduction}. All intensities were reconstructed using Version 1.0.4 of the \emph{Dragonfly} software. The virtual powder sum from all the patterns is shown in Fig.~\ref{fig:powder_mask}(a). Figure~\ref{fig:powder_mask}(b) shows the mask used when reconstructing the intensities. The innermost pixels inside the central speckle were not used to determine the orientations because of saturation. Some other regions were completely excluded from either orientation determination or to calculate the average 3D intensity. 

\begin{figure}[H]
\begin{lstlisting}[language=ini]
[parameters]
detd = 586
lambda = 7.75
detsize = 260 257
pixsize = 0.3
stoprad = 40
ewald_rad = 650.
polarization = x

[make_detector]
in_mask_file = aux/mask_pnccd_back_260_257.byt
out_detector_file = data/det_pnccd_back.dat

[emc]
in_photons_list = amo86615_PR772.txt
in_detector_file = make_detector:::out_detector_file
num_div = 10
output_folder = data/
log_file = EMC.log
need_scaling = 1
beta = 0.001
beta\_schedule = 1.41421356 10
\end{lstlisting}
\caption{Configuration file used to perform \emph{Dragonfly} reconstructions with the entire data set of \SI{14772} patterns.}
\label{fig:dragonfly_config}
\end{figure}

First, the photon converted patterns were downloaded as HDF5 files from the CXIDB. Each file contains patterns from a single experimental run. The photons were first converted to the sparse \texttt{.emc} format using the script \texttt{h5toemc.py}. The configuration file used for this reconstruction is shown in Fig.~\ref{fig:dragonfly_config}. The file specified by \texttt{in\_mask\_file} is provided along with the \emph{Dragonfly} source code and is shown in Fig.~\ref{fig:powder_mask}(b). The \texttt{make\_detector.py} utility was used to generate the detector file detailing which voxel was sampled by every pixel. The \texttt{ewald\_rad} parameter sets the $q$-space size of a voxel which is defined to be \texttt{1/lambda/ewald\_rad}. \texttt{amo86615\_PR772.txt} is a text file containing the names of the converted \texttt{emc} files from every run. 100 iterations of the EMC algorithm were performed starting from a random starting model (uniform random numbers at each voxel). 

For all the cases where the data set was split into two halves, the \texttt{selection} option was added in the \texttt{[emc]} section and set to \texttt{odd\_only} and \texttt{even\_only} for the two halves respectively. Since the intensity reconstruction is invariant to an overall rotation, the two half-data set volumes were rotationally aligned with each other using the \texttt{compare} utility in \emph{Dragonfly}. This program maximizes the overall CC$_{1/2}$ between the two models within a radius range and also calculates the value of CC$_{1/2}$ as a function of $q$ (as shown in Fig.~\ref{fig:split_metrics}(a)).
 
\section*{\label{sec:app_proj}Appendix B: $P_M$ and $P_S$ are projections}
Equation~\ref{eq:pmod} for $P_M$ describes the rescaling of both the background and signal Fourier magnitudes by the square root of the ratio of measured to calculated intensities. The Fourier space modulus constraint requires that the calculated intensity defined in Eq.~\ref{eq:icalc} equals the measured modulus $\sqrt{I_\text{meas}}$. $I_\text{calc}$ has three components at each voxel, namely the real and imaginary parts of the Fourier transform of the electron density and the background, which is allowed to vary independently. The constraint set, therefore, represents the surface of a sphere with radius equalling the measured modulus. The projection of a general point, $\{\operatorname{Re}(\mathcal{F}[\rho]), \operatorname{Im}(\mathcal{F}[\rho]), B\}$ to this sphere is just a rescaling of this 3-vector by the ratio of the magnitudes.

The support projection applies different operations to the two halves of the iterate. For the electron density $\rho(\mathbf{x})$, the ``voxel number" constraint states that at most $N$ voxels have non-zero density. The projection to this constraint set under a Euclidean metric is just to let these $N$ voxels be the ones with the highest absolute value. Note, however, that unlike the conventional fixed support constraint, this ``voxel number'' constraint on $\rho$ is non-convex. For the background volume, $B(\mathbf{q})$, the constraint requires that the background be spherically symmetric. Stated another way, the voxels within the same radial bin should have the same value. The projection to this set is to replace the background magnitude by its azimuthally averaged value. 

\section*{\label{sec:app_masking}Appendix C: Iterative phasing details}
This appendix contains some additional implementation details about the phase retrieval procedure described in Section~\ref{sec:recon_steps}. The code used to perform the reconstructions in this work can be found here: \texttt{https://github.com/andyofmelbourne/3D-Phasing}. The configuration file used is described in Fig.~\ref{fig:phasing_config}. 

\begin{figure}[H]
\centering
\begin{subfigure}[t]{0.43\textwidth}
\begin{lstlisting}[language=ini]
[input]
script = 'make_input.py'
fname = ''
dtype = float
shape = 125, 125, 125
padd_to_pow2 = True
inner_mask = 6
outer_mask = 57
outer_outer_mask = 64
subtract_percentile = None
mask_edges = True
spherical_support = None

[geom]
energy = 2.56348259328e-16
detector_distance = 586.0e-3
voxel_size = 901.538461538e-6
\end{lstlisting}
\end{subfigure}%
\hspace{0.1\textwidth}
\begin{subfigure}[t]{0.43\textwidth}
\begin{lstlisting}[language=ini,escapechar=!]
[phasing]
script = phase.py
repeats = 400
iters = 100ERA 200DM 200ERA

[phasing_parameters]
voxel_number = 2000
support = None
background = True

hardware = cpu
dtype = double

[output]
path = ''

!$\,$!  
\end{lstlisting}
\end{subfigure}
\caption{Configuration file used to reconstruct electron density from the 3D intensity distribution for all data sets.}
\label{fig:phasing_config}
\end{figure}

As in the intensity reconstructions, the central speckle intensities were not found to be trustworthy and were masked out up to a radius of 6 voxels from the center. This means that during the modulus projection $P_M$, these voxels were left unmodified. In addition to this central region, a 7-voxel thick shell at the edge of the sphere of reconstructed intensities was also masked out in order to avoid ringing artifacts due to truncating half a speckle.

As mentioned in Section~\ref{sec:phasing_steps}, the reconstruction from the different random starting guesses need to be aligned with respect to each other before averaging and calculating the PRTF. This is done in three steps, first by translating the volumes such that the center of mass of each of them is at the origin. Second, since the objects are assumed to be complex-valued in general, a global phase is removed by subtracting the mean phase over all voxels. Finally, in order to remove a central inversion uncertainty, one solution (for convenience, the first) is taken as the reference For each of the other solutions, the error with respect to the reference for both the original and the center-inverted version is calculated and the one with lower error is retained.

\section*{Funding}
US Department of Energy (DOE), Office of Science, Office of Basic Energy Sciences (OBES), under contract DE-AC02-76SF00515;
U.S. National Science Foundation (NSF) Science and Technology Center BioXFEL Award 1231306; 
Australian Research Council Centre of Excellence in Advanced Molecular Imaging (AMI);
European Research Council, ``Frontiers in Attosecond X-ray Science: Imaging and Spectroscopy (AXSIS)'', ERC-2013-SyG 609920 (2014-2018);
The Human Frontier Science Program (RGP0010/2017);
Fellowship from the Joachim Herz Stiftung;
Cluster of Excellence ``The Hamburg Center for Ultrafast Imaging'' of the Deutsche Forschungsgemeinschaft (DFG) - EXC 1074 - project ID 194651731;
Helmholtz Association through project-oriented funds.

\section*{Acknowledgments}
We wish to thank the members of the Single Particle Imaging initiative at LCLS who provided valuable feedback regarding this work, such as Ivan Vartanyants, John Spence and Max Rose.

\section*{Disclosures}
The authors declare no conflict of interest.

\bibliography{refs}

\begin{thebibliography}{10}
\newcommand{\enquote}[1]{``#1''}

\bibitem{Neutze:2000}
R.~Neutze, R.~Wouts, D.~van~der Spoel, E.~Weckert, and J.~Hajdu,
  \enquote{{Potential for biomolecular imaging with femtosecond X-ray pulses},}
  {\protect\JournalTitle{Nature}}  (2000).

\bibitem{Ayyer:2016}
K.~Ayyer, T.-Y. Lan, V.~Elser, and N.~D. Loh, \enquote{Dragonfly: an
  implementation of the expand--maximize--compress algorithm for
  single-particle imaging,} {\protect\JournalTitle{Journal of applied
  crystallography}} \textbf{49}, 1320--1335 (2016).

\bibitem{Loh:2010}
N.~D. Loh, M.~J. Bogan, V.~Elser, A.~Barty, S.~Boutet, S.~Bajt, J.~Hajdu,
  T.~Ekeberg, F.~R. N.~C. Maia, J.~Schulz, M.~M. Seibert, B.~Iwan, N.~Timneanu,
  S.~Marchesini, I.~Schlichting, R.~L. Shoeman, L.~Lomb, M.~Frank, M.~Liang,
  and H.~N. Chapman, \enquote{Cryptotomography: Reconstructing 3d fourier
  intensities from randomly oriented single-shot diffraction patterns,}
  {\protect\JournalTitle{Phys. Rev. Lett.}} \textbf{104}, 225501 (2010).

\bibitem{Kassemeyer:2013}
S.~Kassemeyer, A.~Jafarpour, L.~Lomb, J.~Steinbrener, A.~V. Martin, and
  I.~Schlichting, \enquote{Optimal mapping of x-ray laser diffraction patterns
  into three dimensions using routing algorithms,}
  {\protect\JournalTitle{Physical Review E}} \textbf{88}, 042710 (2013).

\bibitem{Ekeberg:2015}
T.~Ekeberg, M.~Svenda, C.~Abergel, F.~R. N.~C. Maia, V.~Seltzer, J.-M.
  Claverie, M.~Hantke, O.~J\"onsson, C.~Nettelblad, G.~van~der Schot, M.~Liang,
  D.~P. DePonte, A.~Barty, M.~M. Seibert, B.~Iwan, I.~Andersson, N.~D. Loh,
  A.~V. Martin, H.~Chapman, C.~Bostedt, J.~D. Bozek, K.~R. Ferguson,
  J.~Krzywinski, S.~W. Epp, D.~Rolles, A.~Rudenko, R.~Hartmann, N.~Kimmel, and
  J.~Hajdu, \enquote{Three-dimensional reconstruction of the giant mimivirus
  particle with an x-ray free-electron laser,} {\protect\JournalTitle{Phys.
  Rev. Lett.}} \textbf{114}, 098102 (2015).

\bibitem{Aquila:2018}
A.~Aquila and A.~Barty, \enquote{{Single Molecule Imaging Using X-ray Free
  Electron Lasers},} in \emph{X-ray Free Electron Lasers,}  (Springer, 2018).

\bibitem{Aquila:2015}
A.~Aquila, A.~Barty, C.~Bostedt, S.~Boutet, G.~Carini, D.~DePonte, P.~Drell,
  S.~Doniach, K.~Downing, T.~Earnest, H.~Elmlund, V.~Elser, M.~Gühr, J.~Hajdu,
  J.~Hastings, S.~Hau-Riege, Z.~Huang, E.~Lattman, F.~Maia, S.~Marchesini,
  A.~Ourmazd, C.~Pellegrini, R.~Santra, I.~Schlichting, C.~Schroer, J.~Spence,
  I.~Vartanyants, S.~Wakatsuki, W.~Weis, and G.~Williams, \enquote{The linac
  coherent light source single particle imaging road map,}
  {\protect\JournalTitle{Structural Dynamics}} \textbf{2}, 041701 (2015).

\bibitem{Munke:2016}
A.~Munke, J.~Andreasson, A.~Aquila, S.~Awel, K.~Ayyer, A.~Barty, R.~J. Bean,
  P.~Berntsen, J.~Bielecki, S.~Boutet, M.~Bucher, H.~N. Chapman, B.~J. Daurer,
  H.~DeMirci, V.~Elser, P.~Fromme, J.~Hajdu, M.~F. Hantke, A.~Higashiura, B.~G.
  Hogue, A.~Hosseinizadeh, Y.~Kim, R.~A. Kirian, H.~K.~N. Reddy, T.-Y. Lan,
  D.~S.~D. Larsson, H.~Liu, N.~D. Loh, F.~R. N.~C. Maia, A.~P. Mancuso,
  K.~M{\"u}hlig, A.~Nakagawa, D.~Nam, G.~Nelson, C.~Nettelblad, K.~Okamoto,
  A.~Ourmazd, M.~Rose, G.~van~der Schot, P.~Schwander, M.~M. Seibert, J.~A.
  Sellberg, R.~G. Sierra, C.~Song, M.~Svenda, N.~Timneanu, I.~A. Vartanyants,
  D.~Westphal, M.~O. Wiedorn, G.~J. Williams, P.~L. Xavier, C.~H. Yoon, and
  J.~Zook, \enquote{Coherent diffraction of single rice dwarf virus particles
  using hard x-rays at the linac coherent light source,}
  {\protect\JournalTitle{Scientific Data}} \textbf{3}, 160064 (2016). Data
  Descriptor.

\bibitem{Reddy:2017}
H.~K. Reddy, C.~H. Yoon, A.~Aquila, S.~Awel, K.~Ayyer, A.~Barty, P.~Berntsen,
  J.~Bielecki, S.~Bobkov, M.~Bucher, G.~A. Carini, S.~Carron, C.~Henry,
  D.~Benedikt, H.~DeMirci, T.~Ekeberg, P.~Fromme, J.~Hajdu, M.~F. Hantke,
  P.~Hart, B.~G. Hogue, A.~Hosseinizadeh, Y.~Kim, R.~A. Kirian, R.~P. Kurta,
  D.~S. Larsson, N.~Loh, F.~R. N.~C. Maia, A.~P. Mancuso, K.~Mühlig, A.~Munke,
  D.~Nam, C.~Nettelblad, A.~Ourmazd, M.~Rose, P.~Schwander, M.~Seibert, J.~A.
  Sellberg, C.~Song, J.~C. Spence, M.~Svenda, G.~Van~der Schot, I.~A.
  Vartanyants, G.~J. Williams, and P.~Xavier, \enquote{Coherent soft x-ray
  diffraction imaging of coliphage pr772 at the linac coherent light source,}
  {\protect\JournalTitle{Scientific Data}} \textbf{4} (2017).

\bibitem{Loh:2009}
N.-T.~D. Loh and V.~Elser, \enquote{Reconstruction algorithm for
  single-particle diffraction imaging experiments,}
  {\protect\JournalTitle{Physical Review E}} \textbf{80}, 026705 (2009).

\bibitem{Ferguson:2015}
K.~R. Ferguson, M.~Bucher, J.~D. Bozek, S.~Carron, J.-C. Castagna, R.~Coffee,
  G.~I. Curiel, M.~Holmes, J.~Krzywinski, M.~Messerschmidt, M.~Minitti,
  A.~Mitra, S.~Moeller, P.~Noonan, T.~Osipov, S.~Schorb, M.~Swiggers,
  A.~Wallace, J.~Yin, and C.~Bostedt, \enquote{The atomic, molecular and
  optical science instrument at the linac coherent light source,}
  {\protect\JournalTitle{Journal of synchrotron radiation}} \textbf{22},
  492--497 (2015).

\bibitem{Maia:2012}
F.~R. Maia, \enquote{The coherent x-ray imaging data bank,}
  {\protect\JournalTitle{Nature methods}} \textbf{9}, 854--855 (2012).

\bibitem{Yoon:2011}
C.~H. Yoon, P.~Schwander, C.~Abergel, I.~Andersson, J.~Andreasson, A.~Aquila,
  S.~Bajt, M.~Barthelmess, A.~Barty, M.~J. Bogan, C.~Bostedt, J.~Bozek, H.~N.
  Chapman, J.-M. Claverie, N.~Coppola, D.~P. DePonte, T.~Ekeberg, S.~W. Epp,
  B.~Erk, H.~Fleckenstein, L.~Foucar, H.~Graafsma, L.~Gumprecht, J.~Hajdu,
  C.~Y. Hampton, A.~Hartmann, E.~Hartmann, R.~Hartmann, G.~Hauser,
  H.~Hirsemann, P.~Holl, S.~Kassemeyer, N.~Kimmel, M.~Kiskinova, M.~Liang,
  N.-T.~D. Loh, L.~Lomb, F.~R. N.~C. Maia, A.~V. Martin, K.~Nass, E.~Pedersoli,
  C.~Reich, D.~Rolles, B.~Rudek, A.~Rudenko, I.~Schlichting, J.~Schulz,
  M.~Seibert, V.~Seltzer, R.~L. Shoeman, R.~G. Sierra, H.~Soltau, D.~Starodub,
  J.~Steinbrener, G.~Stier, L.~Str\"{u}der, M.~Svenda, J.~Ullrich,
  G.~Weidenspointner, T.~A. White, C.~Wunderer, and A.~Ourmazd,
  \enquote{Unsupervised classification of single-particle x-ray diffraction
  snapshots by spectral clustering,} {\protect\JournalTitle{Opt. Express}}
  \textbf{19}, 16542--16549 (2011).

\bibitem{Elser:2003}
V.~Elser, \enquote{Phase retrieval by iterated projections,}
  {\protect\JournalTitle{JOSA A}} \textbf{20}, 40--55 (2003).

\bibitem{Fienup:1978}
J.~R. Fienup, \enquote{Reconstruction of an object from the modulus of its
  fourier transform,} {\protect\JournalTitle{Optics letters}} \textbf{3},
  27--29 (1978).

\bibitem{Kurta:2017}
R.~P. Kurta, J.~J. Donatelli, C.~H. Yoon, P.~Berntsen, J.~Bielecki, B.~J.
  Daurer, H.~DeMirci, P.~Fromme, M.~F. Hantke, F.~R. N.~C. Maia, A.~Munke,
  C.~Nettelblad, K.~Pande, H.~K.~N. Reddy, J.~A. Sellberg, R.~G. Sierra,
  M.~Svenda, G.~van~der Schot, I.~A. Vartanyants, G.~J. Williams, P.~L. Xavier,
  A.~Aquila, P.~H. Zwart, and A.~P. Mancuso, \enquote{Correlations in scattered
  x-ray laser pulses reveal nanoscale structural features of viruses,}
  {\protect\JournalTitle{Phys. Rev. Lett.}} \textbf{119}, 158102 (2017).

\bibitem{Rose:2018}
M.~Rose, S.~Bobkov, K.~Ayyer, R.~P. Kurta, D.~Dzhigaev, Y.~Y. Kim, A.~J.
  Morgan, C.~H. Yoon, D.~Westphal, J.~Bielecki, J.~A. Sellberg, G.~Williams,
  F.~R. Maia, O.~M. Yefanov, V.~Ilyin, A.~P. Mancuso, H.~N. Chapman, B.~G.
  Hogue, A.~Aquila, A.~Barty, and I.~A. Vartanyants, \enquote{Single-particle
  imaging without symmetry constraints at an x-ray free-electron laser,}
  {\protect\JournalTitle{IUCrJ}} \textbf{5} (2018).

\bibitem{Henderson:2012}
R.~Henderson, A.~Sali, M.~L. Baker, B.~Carragher, B.~Devkota, K.~H. Downing,
  E.~H. Egelman, Z.~Feng, J.~Frank, N.~Grigorieff, W.~Jiang, S.~J. Ludtke,
  O.~Medalia, P.~A. Penczek, P.~B. Rosenthal, M.~G. Rossmann, M.~F. Schmid,
  G.~F. Schr{\"o}der, A.~C. Steven, D.~L. Stokes, J.~D. Westbrook, W.~Wriggers,
  H.~Yang, J.~Young, H.~M. Berman, W.~Chiu, G.~J. Kleywegt, and C.~L. Lawson,
  \enquote{Outcome of the first electron microscopy validation task force
  meeting,} {\protect\JournalTitle{Structure}} \textbf{20}, 205 -- 214 (2012).

\bibitem{Shapiro:2005}
D.~Shapiro, P.~Thibault, T.~Beetz, V.~Elser, M.~Howells, C.~Jacobsen, J.~Kirz,
  E.~Lima, H.~Miao, A.~M. Neiman, and D.~Sayre, \enquote{Biological imaging by
  soft x-ray diffraction microscopy,} {\protect\JournalTitle{Proceedings of the
  National Academy of Sciences}} \textbf{102}, 15343--15346 (2005).

\bibitem{Marchesini:2006}
S.~Marchesini, H.~N. Chapman, A.~Barty, M.~R. Howells, J.~H. Spence, C.~Cui,
  U.~Weierstall, and A.~M. Minor, \enquote{Phase aberrations in diffraction
  microscopy,} {\protect\JournalTitle{IPAP Conf. Series}} \textbf{7}, 380--382
  (2006).

\bibitem{VanHeel:2005}
M.~Van~Heel and M.~Schatz, \enquote{Fourier shell correlation threshold
  criteria,} {\protect\JournalTitle{Journal of structural biology}}
  \textbf{151}, 250--262 (2005).

\bibitem{Philipp:2012}
H.~T. Philipp, K.~Ayyer, M.~W. Tate, V.~Elser, and S.~M. Gruner,
  \enquote{Solving structure with sparse, randomly-oriented x-ray data,}
  {\protect\JournalTitle{Optics express}} \textbf{20}, 13129--13137 (2012).

\bibitem{AyyerP:2015}
K.~Ayyer, H.~T. Philipp, M.~W. Tate, J.~L. Wierman, V.~Elser, and S.~M. Gruner,
  \enquote{Determination of crystallographic intensities from sparse data,}
  {\protect\JournalTitle{IUCrJ}} \textbf{2}, 29--34 (2015).

\end{thebibliography}

\end{document}